\documentclass[
twocolumn,
twocolappendix,
]{aastex631}
\usepackage{subfigure}
\usepackage{amsmath}
\usepackage{txfonts}
\usepackage[utf8]{inputenc}
\shorttitle{Conformal Constraint on Sound Speed on the Radius of PSR J0952\textendash{0607}}
\shortauthors{El~Hanafy \& Awad}
\graphicspath{{./}{Figs/}}

\begin{document}

\title{Implications of the Conformal Constraint on Sound Speed on the Radius of PSR J0952\textendash{0607} within Rastall Gravity}

\correspondingauthor{Waleed El Hanafy}
\email{waleed.elhanafy@bue.edu.eg}

\author[0000-0002-0097-6412]{Waleed El Hanafy}
\affiliation{Centre for Theoretical Physics, The British University in Egypt, P.O. Box 43, El Sherouk City, Cairo 11837, Egypt}
\author[0000-0001-8454-5865]{Adel Awad}
\email{awad.adel@gmail.com}
\affiliation{Department of Physics, Faculty of Science, Ain Shams University, Cairo 11566, Egypt}
\affiliation{Centre for Theoretical Physics, The British University in Egypt, P.O. Box 43, El Sherouk City, Cairo 11837, Egypt}



\begin{abstract}
It has been shown that the nonminimal coupling between geometry and matter can provide models for massive compact stars that are consistent with the conformal bound on the sound speed, $0\leqslant {c}_{s}^{2}\leqslant {c}^{2}/3$, where the core density approaches a few times the nuclear saturation density. We impose the conformal upper bound on the sound speed on Rastall's field equations of gravity, with Krori-Barua potentials in the presence of an anisotropic fluid as a matter source, to estimate the radius of the most massive pulsar ever observed, PSR J0952\textendash{0607}. For its measured mass $M = 2.35\pm 0.17\, M_\odot$, we obtain a radius $R=14.087 \pm 1.0186$~km as inferred by the model. We investigate a possible connection between Rastall gravity and the MIT bag model with an equation of state, ${p}_{r}(\rho )\approx {c}_{s}^{2}\left(\rho -{\rho }_{{\rm{s}}}\right)$, in the radial direction, with ${c}_{s}=c/\sqrt{3}$ and a surface density $\rho_\text{s}$ slightly above the nuclear saturation density $\rho_\text{nuc}=2.7\times 10^{14}$~g cm$^{-3}$. The corresponding mass\textendash{radius} diagram is in agreement with our estimated value of the radius and with astrophysical observations of other pulsars at 68\% confidence level.
\end{abstract}



\section{Introduction}
\label{sec:intro}

Neutron Stars (NSs) provide a unique laboratory to prob the properties of matter at high densities and compactness at Black Hole (BH) lower limits. Recent mass measurement, using multicolor light curves, of the companion of the Black\textendash{Widow} (BW) binary pulsars PSR J0952\textendash{0607}, produces $M = 2.35\pm 0.17\, M_\odot$, by \citet{Romani:2022jhd}. This makes it the most massive pulsar ever observed, which exceeds the lower limit on $M_\text{max}$ of white dwarf–pulsar binaries by $\simeq 0.15 M_\odot$ as implied by radio Shapiro-delay techniques. In addition, observational constraints on radius measurements, consequently the mass\textendash{radius} relation, put astrophysical bounds on the Equation of State (EoS)--in which pressure is given as a function of density $p(\rho)$--of matter within compact objects. In this sense, the size of the PSR J0952\textendash{0607} is an important piece of information to understand the physics relevant to such dense matters. This is the aim of the present study. Recent works that address the structure of the pulsar PSR J0952\textendash{0607} are given by \cite{Ecker:2022dlg, Tsaloukidis:2022rus}.

The remarkable progress in astrophysical observations regarding how much squeezable is the NS core raises new questions. The relativistic Shapiro time delay Observations of Millisecond Pulsars (MSPs) measured the mass of the NS PSR J0740+6620 $M= 2.08 \pm 0.07 M_\odot$ \citep{NANOGrav:2019jur,Fonseca:2021wxt}, while  Neutron Star Interior Composition Explorer (NICER) and X-ray Multi-Mirror (XMM) Newton combined data sets have been used to determine its radius $R= 13.7_{-1.5}^{+2.6}$ km \citep{Miller:2021qha} and independently by \cite{Riley:2021pdl}, $R=12.39_{-0.98}^{+1.30}$ km (68\% credible level). Another mass\textendash{radius} analyses of this pulsar using NICER+XMM based on Gaussian process applying a nonparameteric EoS approach estimates its mass $M=2.07 \pm 0.11 M_\odot$ and radius $R=12.34^{+1.89}_{-1.67}$ km \citep{Legred:2021hdx} which is in agreement with \cite{Landry:2020vaw} at 68\% credible level. On the other hand, NICER analyses of the MSP PSR J0030+0451 measured its mass $M=1.44^{+0.15}_{-0.14} M_\odot$ and radius $R= 13.02_{-1.06}^{+1.24}$ km \citep{Miller:2019cac}, and independently by \cite{Raaijmakers:2019qny} they give mass $M= 1.34^{+0.15}_{-.16} M_\odot$ and radius $R= 12.71^{+1.14}_{-1.19}$ km. The PSR J0437\textendash{4715} has a measured mass $M=1.44 \pm 0.07 M_\odot$ \citep{Reardon:2015kba} and radius $R=13.6 \pm 0.9$ km \citep{Gonzalez-Caniulef:2019wzi} by the analyses of the surface X-ray thermal emission. Additionally, the Laser Interferometer Gravitational-Wave Observatory (LIGO)  and Virgo constraints on the radius of a canonical NS using GW170817+GW190814 signals with mass $M=1.4 M_\odot$ determines a radius $R=12.9\pm 0.8$ km \citep{LIGOScientific:2020zkf}. Although NICER observations of the NS PSR J0740+6020 indicate much more mass than PSR J0030+0451 or LIGO canonical NS, they have almost the same size $R\simeq 13$ km. The result represents a challenge to both more and less squeezable theoretical models, since more squeezable models, whereas neutrons are suggested to be squeezed to quarks in the NS core, seem to be disfavored by NICER observations of NSs radii. On the other hand, less squeezable models suggest a rapid increase of the pressure at the NS core to support the existence of massive NSs with masses $\simeq 2 M_\odot$, which requires a problematic nonmonotonical rapid increase of the sound speed strongly violating the conjectured conformal upper bound $c_s=c/\sqrt{3}$ somewhere inside the NS \citep{McLerran:2018hbz}, see also the recent study by \cite{Altiparmak:2022bke}.

The Gravitational Wave (GW) observations of NS\textendash{NS} merging, GW170817 \citep{TheLIGOScientific:2017qsa,LIGOScientific:2018cki} and GW190425 \citep{LIGOScientific:2020aai}, indicate no tidal deformability in gravitational wave signals which yields to an upper bound on the stiffness of the EoS at densities relevant to the nuclear saturation density $\rho_\text{nuc}\approx 2.7 \times 10^{14}$ g/cm$^3$ (equivalently the baryon bensity number $n_\text{nuc}\approx 0.16~\text{fm}^{-3}$) in line with microscopic calculations of nuclear matter. On the contrary, existence of massive NSs with masses $\simeq 2M_\odot$ suggests a rapid increase of the pressure of the matter (hadronic/quarkyonic) at densities few times $\rho_\text{nuc}$ at the NS core. This requires a sound speed, $c_s^2= dp/d\rho$, with a large fractional of the speed of light $c$ at the NS core \citep{Alsing:2017bbc,Reed:2019ezm,Landry:2020vaw,Legred:2021hdx,Miller:2021qha}; and then a relatively high stiff EoS which is in tension with small tidal deformability of the GW.

\begin{figure}[ht!]
\includegraphics[width=0.45\textwidth]{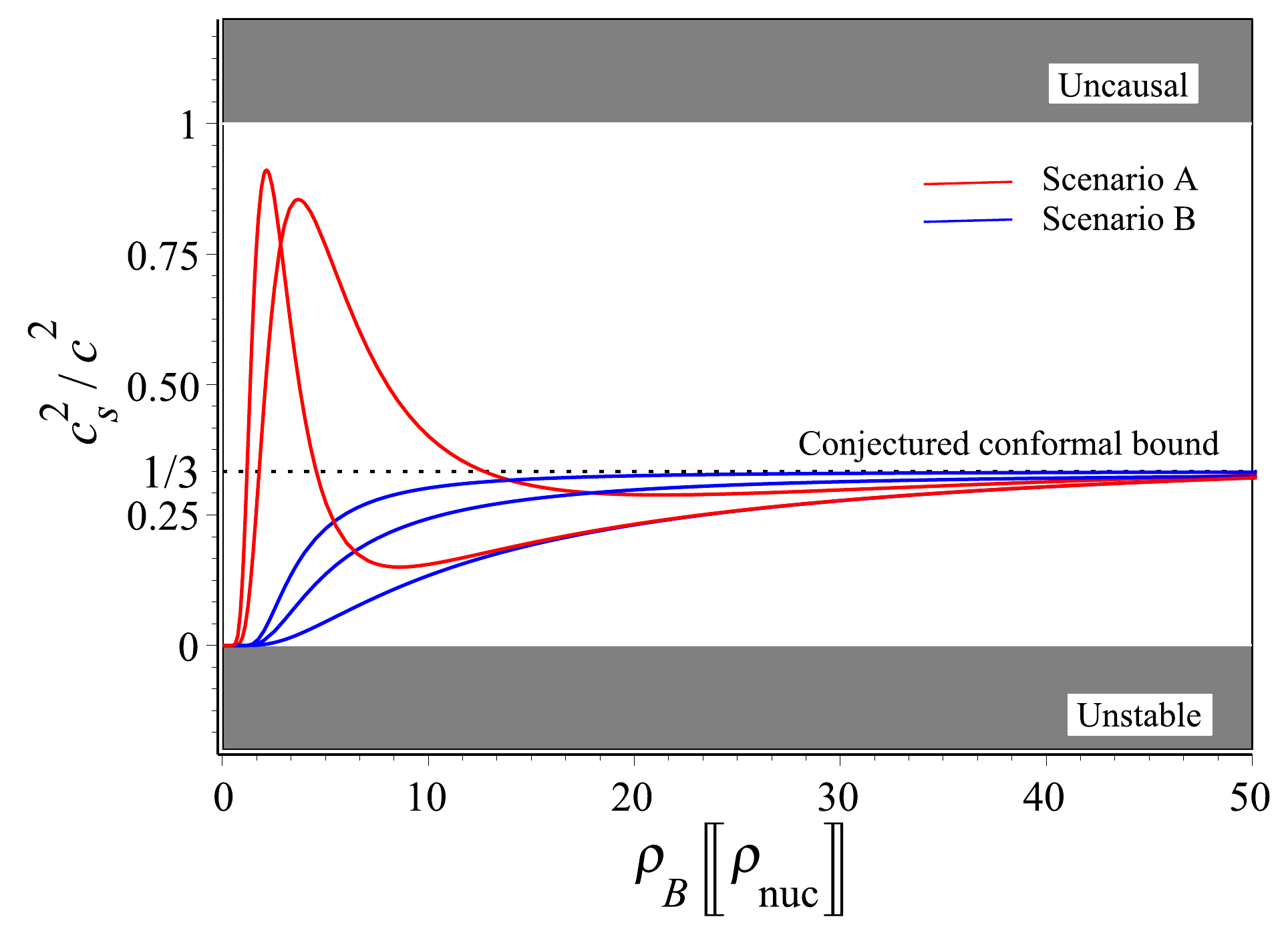}
\caption{Schematic plot to show two possible scenarios as suggested for the sound speed to evolve with baryon matter density, $\rho_B$, given in terms of the nuclear saturation density. Scenario A, the speed of sound nonmonotonically evolves with density reaching a maximum above the conjectured conformal upper bound, $c^2=c^2/3$, at density $\simeq 3-5 \,\rho_\text{nuc}$, and then decreases to approach the conformal bound from below at higher densities $\simeq 40 \, \rho_\text{nuc}$ where perturbative QCD is applied. Scenario B, the speed of sound evolves monotonically with density obeying the conjectured conformal upper bound at all densities. The shaded regions represent the excluded values of the sound speed as constrained by the stability and causality conditions.
\label{Fig:Scenarios}}
\end{figure}

The adiabatic sound speed provides a direct measure of the physical features of matter within the compact object and strongly related to its EoS. In general, for physical matter to be causal and thermodynamically stable it is required that $0 \leq c_s^2/c^2 \leq 1$. However, one finds a more restrictive range whereas non-relativistic particles acquire $c_s^2/c^2 \ll 1$, ultra-relativistic (massless) particles acquire $c_s^2/c^2=1/3$ and the interaction between particles at extremely high densities $\simeq 40 \, \rho_\text{nuc}$, where the Quantum-Chromodynamics (QCD) perturbative approach (i.e. weak coupling) is applied, yields to an upper bound $c_s^2/c^2 \to 1/3$. Between these two limits, at densities a few times $\rho_\text{nuc}$, two possible scenarios are suggested as indicated by Figure \ref{Fig:Scenarios} \citep[see][]{Bedaque:2014sqa,Tews:2018kmu,Moustakidis:2016sab}. Scenario A, the speed of sound nonmonotonically evolves with density reaching a maximum above the conjectured conformal upper bound at density $\simeq 3-5 \,\rho_\text{nuc}$, and then decreases to approach the conformal bound from below at higher densities $\simeq 40 \, \rho_\text{nuc}$ \citep{McLerran:2018hbz,Drischler:2020fvz,Drischler:2021bup}. Scenario B, the speed of sound evolves monotonically with density obeying the conjectured conformal upper bound at all densities \citep{Cherman:2009tw,Bedaque:2014sqa}. It has been shown that nonminimal coupling between matter and geometry, unlike General Relativity (GR), provides a physically viable framework processing scenario B even in the case of the most massive NS and in agreement with astrophysical observations \citep{ElHanafy:2022kjl}. Also, the so-called two-families scenario has been suggested to cope with the sound speed conformal limit as investigated by \cite{Traversi:2021fad} \citep[see also][]{Drago:2013fsa,Drago:2015cea,DePietri:2019khb}.

In this study, following the recent treatment of the pulsar PSR J0740$+$6620 \citep{ElHanafy:2022kjl}, we apply Rastall theory (RT) of gravity, which assumes matter\textendash{geometry} coupling, using scenario B for the BW pulsar PSR J0952\textendash{0607} to constrain its radius. This assumes that the matter inside the NS to be anisotropic fluid which is a natural choice in the presence of high pressures as expected in an NS core \citep{herrera1997local}. Consequently, for the cases where the tangential pressure is greater than the radial, one expects a repulsive anisotropic force to partly relax the attractive gravitational force, as predicted by Tolman–Oppenheimer–Volkoff (TOV) equation \citep{bowers1974anisotropic}, holding the size of massive NSs with no collapse to smaller sizes. Thus, we organize the present work as follows: In Section \ref{Sec:Model}, we review Rastall's field equations of gravity of Krori \& Barua model in presence of anisotropic fluid with the boundary conditions which fix the model parameters. In Section \ref{Sec:Radius}, we investigate the consequences of scenario B on the radius measurement of the BW pulsar PSR J0952\textendash{0607}. In Section \ref{Sec:Strange-star}, we further investigate the possibility of this pulsar being a strange star compatible with scenario B. In Section \ref{Sec:Mass-Radius}, we discuss the crustal structure of the BW pulsar. In addition, we investigate the mass\textendash{radius} relation as obtained by the present model, for a boundary density at nuclear saturation density, and its compatibility with several well measured mass\textendash{radius} data. In Section \ref{Sec:Conclusion}, we conclude the work.

\section{Theoretical Framework of the Model}\label{Sec:Model}

In RT, the matter divergence-free energy-momentum tensor, $\mathfrak{T}{^\alpha}{_\beta}$, assumption has been replaced by $\nabla_{\alpha}\mathfrak{T}{^\alpha}{_\beta}=\frac{\epsilon}{\kappa}\, \partial_{\beta} \mathfrak{R} \neq 0$, where $\nabla$ is the covariant derivative associated with Levi-Civita connection, $\mathfrak{R}$ is Ricci scalar, $\epsilon$ and $\kappa$ are two constants \citep{Rastall:1972swe,Rastall:1976uh}. Therefore, the corresponding field equations can be written as
\begin{equation}\label{RT2}
     \mathfrak{G}_{\alpha\beta}=\kappa \widetilde{\mathfrak{T}}_{\alpha\beta},
\end{equation}
where the effective energy-momentum tensor is written in terms of the matter sector as
\begin{equation}\label{RTmn}
    \widetilde{\mathfrak{T}}_{\alpha\beta}=\mathfrak{T}_{\alpha\beta}+\frac{\epsilon}{1-4\epsilon}g_{\alpha\beta}\mathfrak{T}. \qquad (\epsilon\neq \tfrac{1}{4})
\end{equation}
The above equation reflects the nonminimal coupling nature between matter and geometry in RT. The Newtonian limit of Rastall gravity requires a rescaled effective gravitational constant \citep[c.f.][]{Rastall:1972swe,Moradpour:2017ycq,Moradpour:2016ubd}
\begin{equation}\label{Rconst}
    \kappa=\tfrac{1-4\epsilon}{1-6\epsilon}\,\kappa_E , \qquad (\epsilon\neq \tfrac{1}{6})
\end{equation}
where $\kappa$ reduces to Einstein coupling constant $\kappa_E=8 \pi G/c^4$, with $G$ being the Newtonian constant, only if $\epsilon=0$. Clearly, in RT of gravity, the conservation law is locally broken in curvatured spacetime due to matter\textendash{geometry} coupling. Otherwise, in vacuum, the theory reduces to GR. Therefore, any deviation from GR can be quantified in terms of the dimensionless Rastall parameter $\epsilon$ and the coupling constant $\kappa$ via the effective energy-momentum tensor $\widetilde{\mathfrak{T}}_{\alpha\beta}$. Many studies, like the present one, show the inequivalence of RT and GR, \citep[c.f.,][]{Oliveira:2015lka,Moradpour:2017ycq,Li:2019jkv,Lin:2020fue,2020EPJP..135..916Z,Nashed:2022zyi}, in agreement with \citet{Darabi:2017coc} conclusion and contrary to the claim in \citet{Visser:2017gpz}. Clearly, NSs are perfect laboratories to test matter\textendash{geometry} coupling such as in RT whereas matter density and spacetime curvature are maximal for an observed compact object.

For a nonrotating static spherically symmetric NS, we take the 4-dimensional spacetime line element in a spherical polar coordinates ($t,r,\theta,\phi$)
\begin{equation}\label{eq:metric}
    ds^2=-e^{\alpha(r)}c^2 dt^2 + e^{\beta(r)} dr^2+ r^2 (d\theta^2+\sin^2 \theta \, d\phi^2).
\end{equation}
We further assume the metric potentials inside the star are given by \cite{Krori1975ASS} ansatz (hearafter KB),
\begin{equation}\label{eq:KB}
    \alpha(r)=a_0 x^2+a_1,\,  \beta(r)=a_2 x^2,
\end{equation}
where the dimensionless radius $0 \leq x=r/R \leq 1$ with $R$ being the radius of the star. The set of constants \{$a_0, a_1, a_2$\} are dimensionless and can be determined by matching conditions on the boundary surface of the star. At this moment, it is important to investigate the viability of the static assumption as given by the metric \eqref{eq:metric} in the case of a fast spinning pulsar such as PSR J0952\textendash{0607} with a frequency $f=709$ Hz \citep{Romani:2022jhd}. We refer to the quasi-universal relation which expresses the maximum mass of a uniformly rotating star in terms of the dimensionless angular momentum $j:=c J/G M^2$, where $J$ is the total angular momentum of the star \citep{Breu:2016ufb}
\begin{equation}\label{eq:BR_formula}
    \frac{M_\text{crit}}{M_\text{TOV}}=1+\bar{a}_2 \left(\frac{j}{j_\text{Kep}}\right)^2+\bar{a}_4 \left(\frac{j}{j_\text{Kep}}\right)^4,
\end{equation}
where $j_\text{Kep}:=c J_\text{Kep}/G M_\text{Kep}^2$ is the Keplerian dimensionless angular momentum, $M_\text{crit}(j=j_\text{Kep})=M_\text{max}$ and $M_\text{crit}(j=0)=M_\text{TOV}$. The coefficients $\bar{a}_2 = 0.1316$ and $\bar{a}_4 = 0.07111$ as obtained by \cite{Breu:2016ufb}. Noting that, the maximum rotational frequency $f_\text{Kep}\approx 1.45$ kHz, for the soft EoS as estimated by \cite{Demircik:2020jkc} with maximum $c_{s}^2 \approx  0.42$. For the particular case of the pulsar PSR J0952\textendash{0607} with a rotating frequency of $f=709$ Hz, one have $j/j_\text{Kep} \approx f/f_\text{Kep} \lesssim 0.49$ \citep{Ecker:2022dlg}. This gives
\begin{equation}
   \frac{M_\text{crit}}{M_\text{TOV}}=1.0352,
\end{equation}
i.e. the mass increase due to rotational frequency is, $\simeq 3.5\%$ ($\simeq 0.08 M_\odot$), much less than the mass uncertainty, $M=2.35 \pm 0.17 M_\odot$, of the pulsar PSR J0952\textendash{0607}. This justifies well the nonrotating assumption in our treatment, see also \cite{Ecker:2022dlg}. Although the above estimate is based on hadronic EoSs, it has been shown that hybrid matter (including phase transition from hadron to quark) would not alter the above mentioned conclusion \citep{Bozzola:2019tit, Demircik:2020jkc}. Therefore, we do not expect very different results for quark matter, especially, since the changes in the mass\textendash{radius} relation for quark/hardron EoSs are not significant for mass $\simeq 2 M_\odot$, see \citep{Bhattacharyya:2016kte}. Similar analysis could be done to check the viability of the static approximation with the radius measurement. We note that the radius of the BW pulsar PSR J0952\textendash{0607} is not measured yet, and therefore only mass could be relevant in this case. However, we provide more detailed discussion on this issue in appendix \ref{App:Radius_static_approx}.

On the other hand, we, additionally, assume the energy-momentum tensor for a anisotropic fluid with spherical symmetry, i.e.
\begin{equation}\label{Tmn-anisotropy}
    \mathfrak{T}{^\alpha}{_\beta}=(p_{t}+\rho c^2)U{^\alpha} U{_\beta}+p_{t} \delta ^\alpha _\beta + (p_{r}-p_{t}) V{^\alpha} V{_\beta},
\end{equation}
where $U_\alpha$ is the time-like four-velocity and $V{^\alpha}$ is the unit space-like vector in the radial direction; while $\rho$, $p_{r}$ and $p_{t}$ are the fluid energy density, radial pressure ($U_\alpha$-direction) and tangential pressure (perpendicular to $U_\alpha$), respectively. Then, the energy-momentum tensor takes the diagonal form $\mathfrak{T}{^\alpha}{_\beta}=diag(-\rho c^2,\,p_{r},\,p_{t},\,p_{t})$, and similarly the effective energy-momentum tensor reads $\widetilde{\mathfrak{T}}{^\alpha}{_\beta}=diag(-\tilde{\rho} c^2,\,\tilde{p}_{r},\,\tilde{p}_{t},\,\tilde{p}_{t})$ whereas Equation \eqref{RTmn} gives
\begin{eqnarray}\label{eq:Eff-dens-press}
\nonumber \tilde{\rho} c^2 &=& \rho c^2 + \frac{\epsilon}{1-4\epsilon} (\rho c^2 - p_r - 2p_t),\\
\nonumber \tilde{p}_r &=& p_r - \frac{\epsilon}{1-4\epsilon} (\rho c^2 - p_r - 2p_t),\\
\tilde{p}_t &=& p_t - \frac{\epsilon}{1-4\epsilon} (\rho c^2 - p_r - 2p_t).
\end{eqnarray}

Applying Rastall's field equations \eqref{RT2} to the spacetime \eqref{eq:metric} where the matter sector is given by \eqref{Tmn-anisotropy}, we obtain the following dimensionless field equations \citep{ElHanafy:2022kjl}
\begin{eqnarray}
\eta \bar{\rho}&=& \frac{e^{-a_2 x^2}}{x^2}(e^{a_2 x^2}-1+2a_2 x^2) \nonumber\\
&+&\frac{2\epsilon}{x^2}\left[\left(a_0(a_0-a_2)x^4 -(2a_2-3a_0)x^2 +1\right)e^{-a_2x^2} -1\right],\nonumber\\[8pt]
\eta \bar{p}_r&=&\frac{e^{-a_2 x^2}}{x^2}(1-e^{a_2 x^2}+2a_0 x^2) \nonumber\\
&-&\frac{2\epsilon}{x^2}\left[\left(a_0 (a_0-a_2) x^4 - (2 a_2-3 a_0) x^2 +1\right)e^{-a_2 x^2} - 1\right],\nonumber\\[8pt]
\eta \bar{p}_t&=& e^{-a_2 x^2}(2 a_0-a_2 +a_0 (a_0 - a_2) x^2) \nonumber\\
&-&\frac{2\epsilon}{x^2}\left[\left(a_0 (a_0-a_2) x^4 - (2 a_2-3 a_0) x^2 +1\right)e^{-a_2 x^2} - 1\right],\nonumber \\
\label{eq:Feqs2}
\end{eqnarray}
where
\begin{equation}
   \eta=\frac{1-4\epsilon}{1-6\epsilon}, \, \bar{\rho}(r)=\frac{\rho(r)}{\rho_{\star}},\, \bar{p}_r(r)=\frac{p_r(r)}{\rho_{\star} c^2},\, \bar{p}_t(r)=\frac{p_t(r)}{\rho_{\star} c^2},
\end{equation}
and $\rho_{\star}$ denotes a characteristic density
\begin{equation}
    \rho_{\star}=\frac{1}{\kappa_E c^2 R^2}.
\end{equation}
Equations \eqref{eq:Feqs2} ensure that the anisotropy, $\Delta:=p_t-p_r$, is null at the center and positive everywhere else, for more details see \citep{ElHanafy:2022kjl}. Using the density and radial pressure \eqref{eq:Feqs2}, we induce an EoS of the matter inside the star ($0\leq x<1$) up to $O(x^4)$
\begin{equation}\label{eq:linear_EoS}
\bar{p}_r(\bar{\rho})\approx c_1 \bar{\rho} + c_0,
\end{equation}
where
\begin{eqnarray}\label{eq:EoS-const}
\nonumber c_1&=& {(4a_0-a_2)a_2-2(8 a_0 a_2 -2a_0^2 -5 a_2^2) \epsilon \over 5 a_2^2 + 2(8 a_0 a_2 -2a_0^2 -5 a_2^2) \epsilon}, \\
c_0&=& -{2(1-6\epsilon)(a_0+a_2)[a_2^2+2(2a_0^2-2a_0 a_2 -a_2^2)\epsilon] \over (1-4\epsilon) [5 a_2^2 + 2(8 a_0 a_2 -2a_0^2 -5 a_2^2) \epsilon]}.\quad
\end{eqnarray}
Clearly, the KB ansatz \eqref{eq:KB}, in general, induces a linear EoS similar to the popular MIT bag model, whereas $c_1$ and $c_0$ represent a dimensionless sound speed squared and bag constant respectively. The EoS \eqref{eq:linear_EoS} can be rewritten as
\begin{equation}\label{eq:MIT-EoS}
    p_r(\rho)=c_s^2(\rho-\rho_\text{s}),
\end{equation}
where $\rho_\text{s}$ is the density at zero pressure (at the surface). By comparison we obtain $c_1=c_s^2/c^2$ and $c_0=-c_1 \left(\rho_\text{s}/\rho_\star \right)$. We note that, for anisotropic star models, one may need to assume two EoSs: One for the radial pressure $p_r(\rho)$ and one for the tangential pressure $p_t(\rho)$ to close the system. In the present approach we instead use KB ansatz \eqref{eq:KB} (which relates the density and pressures). Therefore, in principal, no EoS are assumed or needed. Otherwise the system is over-determined. The induced EoS \eqref{eq:MIT-EoS}, however, is useful when investigating possible matter content and the model viability on the microphysics level as will be discussed later in Section \ref{Sec:Strange-star}.

For vacuum case, GR and RT are equivalent. Therefore, the exterior spacetime is exactly Schwarzschild's vacuum solution
\begin{equation}
    ds^2=-\left(1-\frac{2GM}{c^2r}\right) c^2 dt^2+\frac{dr^2}{\left(1-\frac{2GM}{c^2 r}\right)}+r^2 (d\theta^2+\sin^2 \theta d\phi^2),
\end{equation}
where $M$ is the mass of the compact object. Recalling the interior spacetime \eqref{eq:metric}, we thus take
\begin{equation}\label{eq:bo}
    \alpha(x=1)=\ln(1-C),\, \beta(x=1)=-\ln(1-C),
\end{equation}
where $C=\frac{2GM}{c^2 R}$ is the compactness parameter. In addition, we make the radial pressure to vanish on the boundary, i.e.
\begin{equation}\label{eq:bo2}
    \bar{p}_r(x=1)=0.
\end{equation}
Using KB ansatz \eqref{eq:KB} and the radial pressure \eqref{eq:Feqs2} along with the above boundary conditions, we have \citep{ElHanafy:2022kjl}
\begin{eqnarray}
a_0&=&-\frac{3}{2}-\frac{1}{2}\ln(1-C)-\frac{\sqrt{\chi}+(1-C)}{2\epsilon(1-C)}, \nonumber \\
a_1&=&\frac{3}{2}+\frac{3}{2}\ln(1-C)+\frac{\sqrt{\chi}+(1-C)}{2\epsilon(1-C)}, \nonumber \\
a_2&=&-\ln(1-C),\label{eq:const}
\end{eqnarray}
where
\begin{eqnarray}
    \chi&=&\epsilon^2 (1-C)^2 \ln(1-C)^2 -2\epsilon (1+\epsilon ) (1-C)^2 \ln(1-C) \nonumber\\
    &-&(1-C)[(5C-9)\epsilon^2 -2(2C-3)\epsilon -(1-C)].
\end{eqnarray}
Taking the limit $\epsilon \to 0$, the set of constants \eqref{eq:const} reduces to the GR version \citep{Roupas:2020mvs}
\begin{equation}
    a_0=\frac{C}{2(1-C)},\, a_1=\ln(1-C)-\frac{C}{2(1-C)},\, a_2=-\ln(1-C).
\end{equation}
Interestingly all physical quantities of KB spacetime within a given star, $0\leq x\leq 1$, can be written as dimensionless forms in terms of Rastall and the compactness parameters, i.e. $\bar{\rho}(\epsilon,C)$, $\bar{p}_r(\epsilon,C)$ and $\bar{p}_t(\epsilon,C)$.
\begin{figure*}[ht!]
\centering
\subfigure{\includegraphics[width=0.49\textwidth]{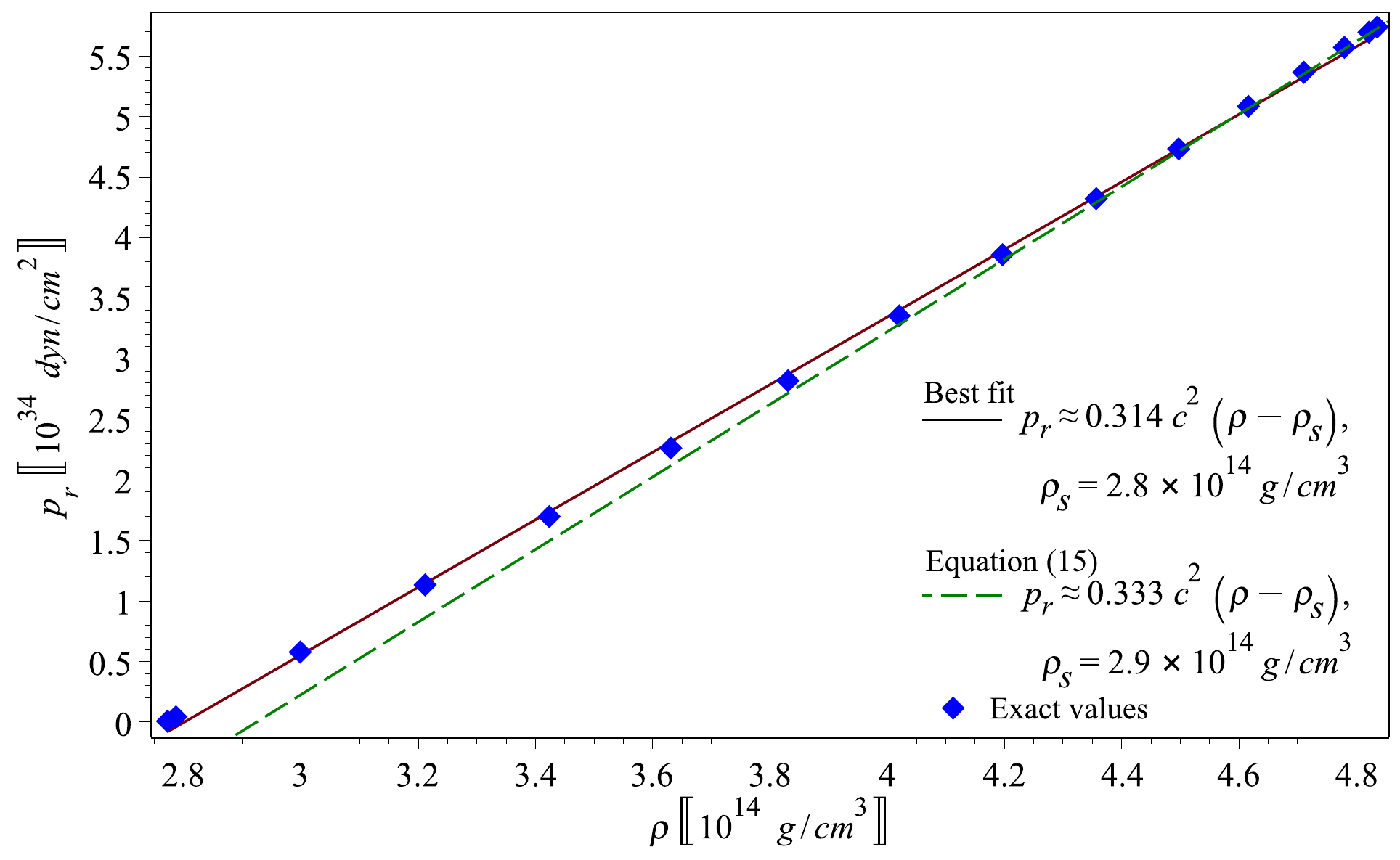}}\hspace{0.2cm}
\subfigure{\includegraphics[width=0.49\textwidth]{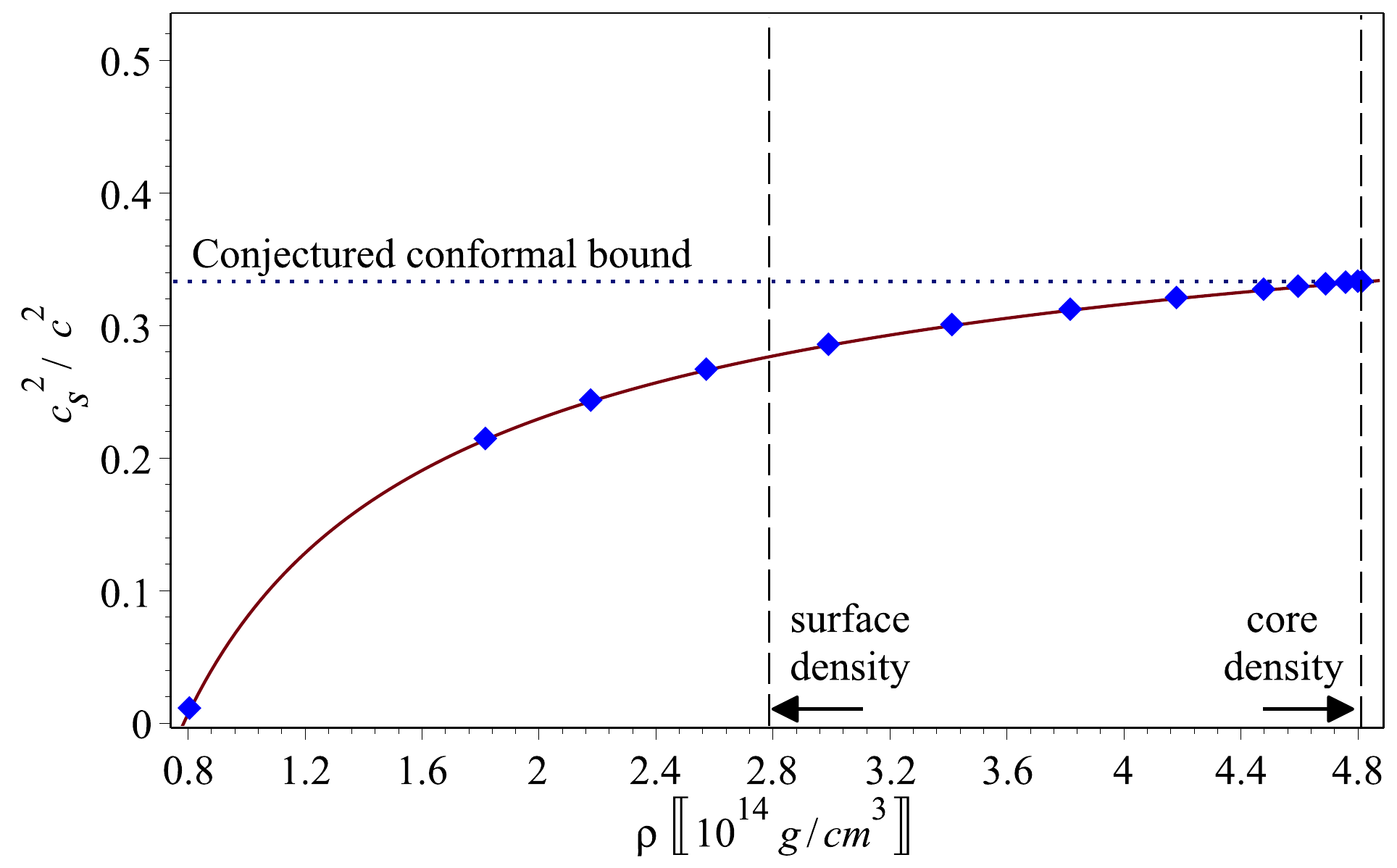}}
\caption{Left panel: The evolution of the radial pressure with the density inside the PSR J0952\textendash{0607}. We use the numerical values as obtained by the model \{$a_0=0.44021$, $a_1=-1.11916$ and $a_2=0.67894$, $\epsilon=0.041$, $C=0.495$, $R=14.087$ km\}. Substituting in the field equations \eqref{eq:Feqs2}, we plot the pressure\textendash{density} diagram. The model perfectly fits with the linear pattern $p_r(\rho)\approx 0.314 c^2 \left(\rho -\rho_\text{s}\right)$ where the surface density $\rho_\text{s}=2.8 \times 10^{14}$~g/cm$^3$ just above $\rho_\text{nuc}$, and the slope $c_s^2\approx 0.314 c^2$ (black solid line). It is in agreement with the induced EoS as obtained by Equation \eqref{eq:MIT-EoS} with a slightly greater surface density $\rho_\text{s}=2.9 \times 10^{14}$~g/cm$^3$ and slope $c_s^2\approx 0.333 c^2$ (green dashed line). Right panel: The evolution of the sound speed squared with the density inside the PSR J0952\textendash{0607}, $\rho \approx (2.8-4.8) \times 10^{14}$~g/cm$^3$ , whereas the conjectured conformal limit is fulfilled everywhere inside the star.
\label{Fig:EoS}}
\end{figure*}
%
%

\section{Sound speed constraint on the radius of PSR J0952\textendash{0607}}\label{Sec:Radius}
It has been shown that combined data of NICER+XMM observations of the pulsar PSR J0740$+$6620 determines Rastall parameter $\epsilon=0.041$ \citep{ElHanafy:2022kjl}, which will be used here to determine the radius of the pulsar PSR J0952\textendash{0607}. Using the induced EoS \eqref{eq:linear_EoS}, we define the square of the sound speed in radial direction,
\begin{equation}\label{eq:sound_speed}
  c_s^2 = \frac{dp_r}{d\rho}= \frac{\bar{p}'_r}{\bar{\rho}'},
\end{equation}
where $'=R^{-1} d/dx$. We require the maximum sound speed at the center to obey the conjectured conformal upper bound $c_s^2(r=0)/c^2 \lesssim 1/3$ where the density at the core is maximal. This sets the following constraint on the field equations \eqref{eq:Feqs2}
\begin{equation}\label{eq:speed_limit}
    {(4a_0-a_2)a_2-2(8 a_0 a_2 -2a_0^2 -5 a_2^2) \epsilon \over 5 a_2^2 + 2(8 a_0 a_2 -2a_0^2 -5 a_2^2) \epsilon} \lesssim \frac{1}{3}.
\end{equation}
Using the constraint $c_1\lesssim 1/3$, in Equations \eqref{eq:EoS-const}, the above inequality is re-obtained. Given that the mass of the pulsar PSR J0952\textendash{0607} is $M = 2.35\pm 0.17\, M_\odot$ \citep{Romani:2022jhd}, and using the set of equations \eqref{eq:const} in addition to the conformal sound speed limit \eqref{eq:speed_limit}, we estimate the radius $R=14.087 \pm 1.0186$ km. This determines the set of constants \{$a_0=0.44021\pm 0.00001$, $a_1=-1.11916\pm 0.00003$ and $a_2=0.67894\pm 0.00001$\}. Consequently, we determine the compactness parameter of PSR J0952\textendash{0607}, $C=0.495 \pm 0.0713$. For Rastall parameter $\epsilon=0.041$, it has been shown that all energy conditions related the effective energy-momentum tensor of Rastall gravity are satisfied if the Dominant Energy Condition (DEC), $\rho c^2-p_r-2 p_t \geq 0$, of the matter sector is fulfilled. On the other hand that DEC sets a restrictive upper limit on the compactness $C=0.735$, which is $2\%$ higher than the GR treatment \citep{ElHanafy:2022kjl}. This confirms that the PSR J0952\textendash{0607} is physically stable.
Notably the BW radius according to our calculations is compatible with $\simeq 2 M_\odot$ pulsars, e.g. PSR J0740+6620, PSR J0348+0432 and PSR J1614–2230, measurements of their radii $R\simeq 13$ km. On the contrary, for the GR limit, $\epsilon \to 0$, and by applying the same approach we estimate the BW radius $R=16.41 \pm 1.19$ km, which is much larger than the measured radii of the pulsars within the relevant mass range. If one wants to keep the BW radius at $R\simeq 13-14$ km, the sound speed then should be $c_s^2/c^2\simeq 0.4-0.37$. This may point out some difficulties of the GR gravity to keep the conformal limit of the sound speed valid while keeping a good agreement with other mass\textendash{radius} measurements. It also has severe consequences on the inviability of the model on the microphysics level as will be clarified latter in Section \ref{Sec:Strange-star}.

We next estimate numerical values of some physical parameters as predicted by Equations \eqref{eq:Feqs2} of the present model: At the core, we calculate the core density $\rho_\text{c}=4.82\times 10^{14}$ g/cm$^3$ which is $\approx 1.8\, \rho_\text{nuc}$, the radial pressure $p_{r,\text{c}}=5.69 \times 10^{34}$ dyn cm$^{-2}$ where the sound speed $c_s^2=0.333 c^2 < c^2/3$. Those are the maximal values as allowed by the model which clearly confirms the consistency with the conjectured conformal upper limit on the sound speed. At the star surface, i.e. $R=14.087$ km as predicted by the model, we calculate the surface density $\rho_\text{s}=2.76\times 10^{14}$ g/cm$^3$ slightly above the saturation nuclear density. Also, the surface redshift $Z_\text{s} := \frac{1}{\sqrt{-g_{tt}}}-1= 0.404$ in agreement with Buchdahl limit \citep{PhysRev.116.1027}.
We plot the radial pressure verses density within PSR J0952\textendash{0607} from core to surface as given by Figure \ref{Fig:EoS} (left panel). We note that the pressure\textendash{density} plot can be obtained directly from the field equations \eqref{eq:Feqs2} as presented by diamond labels. The best fit (black solid line) is highly compatible with linear pattern, $p_r(\rho)\approx 0.314 c^2\left(\rho - \rho_\text{s}\right)$ with $\rho_\text{s}=2.8\times 10^{14}$~g/cm$^3$, similar to MIT bag model. This is, on the other hand, in agreement with the semi-analytical EoS form as given by Equation \eqref{eq:linear_EoS}, where $c_1\simeq 1/3$ and $c_0 \simeq -0.36$ as obtained by Equations \eqref{eq:EoS-const}. Consequently, we plot Equation \eqref{eq:MIT-EoS}, $p_r(\rho)=0.333 c^2\left(\rho - \rho_\text{s}\right)$, with $\rho_\text{s}=-(c_0/c_1)\rho_\star \approx 2.9\times 10^{14}$~g/cm$^3$, as obtained by the green dashed line. Moreover, the evolution of the sound speed at all densities from the surface to the core within the pulsar is given by Figure \ref{Fig:EoS} (right panel). Clearly, the sound speed monotonically increases from the surface (with surface density $\approx \rho_\text{nuc}$) to the core (with core density $\approx 1.8 \,\rho_\text{nuc}$) approaching the conformal upper limit $c_s^2/c^2 \lesssim 1/3$ from below at the core. In the following section we discuss the consequences of the obtained results on the matter content within the BW pulsar.

\section{matter\textendash{geometry} nonminimal coupling mimics strange quark matter}
\label{Sec:Strange-star}

We extend our investigation to examine matter content of the BW pulsar with possible additional constraints on the microphysics level. In this regard, we use the EoS \eqref{eq:MIT-EoS} as induced by KB ansatz \eqref{eq:KB}. By virtue of the imposed constraint \eqref{eq:speed_limit}, as obtained at perturbative QCD limit, it is reasonable to consider the quark matter possibility for non-interacting massless quarks, whereas the EoS in the most simple case is given by the MIT bag model,
\begin{equation}\label{eq:MIT-EoS2}
    p_r=c_s^2(\rho-4B),
\end{equation}
where the speed of sound is fixed to $c_s^2=c^2/3$ and the bag constant $B$ is related to the surface density (at which the radial pressure vanishes) by $\rho_\text{s}=4B$. A pure quark star is possible, if strange quark matter is the true ground state of strongly interacting matter as conjectured by \citet{Witten:1984rs}, see  also \citep{Farhi:1984qu}. This implies a new minimum of the energy per baryon $\rho c^2/n$, at zero pressure, lower than the energy per baryon of Iron nuclei $^{56}$Fe ($ < 930~\text{MeV}$). This makes the quark star a self-bound object, where the surface pressure vanishes while the surface energy density is still large. The divergence of the adiabatic index at the surface characterizes this object and allows for very stable heavy stars $\simeq 2 M_\odot$ even if the sound speed is small. This could provide an evidence for quark matter in massive stars \citep{Annala:2019puf}. Moreover, quark star radius decreases as mass decreases unlike compact stars which are formed from normal neutron matter. Therefore, the mass\textendash{radius} measurements of light compact stars, $M< M_\odot$, could provide another strong evidence on quark star existence \citep{2022NatAs...6.1444D}. The KB model of Rastall gravity as previously studied by \citet{ElHanafy:2022kjl}, and as in the present study, remarkably, shares features similar to quark star models.\\
In order to investigate the similarities between Rastall gravity and the MIT bag model more precisely, we rewrite the field equations \eqref{eq:Eff-dens-press} in the form of
\begin{equation}
    \tilde{\rho}=\rho +\tilde{B}, \, \tilde{p}_r=p_{r}-\tilde{B} c^2, \, \tilde{p}_t=p_{t}-\tilde{B} c^2
\end{equation}
where $\tilde{B}=\frac{\epsilon}{1-4\epsilon}(\rho-p_{r}/c^2-2p_{t}/c^2)$. This leads to the following relation
\begin{equation}\label{eq:eff-EoS}
    \tilde{p}_r=\frac{1}{3}(\tilde{\rho}-4 B_{eff})c^2,
\end{equation}
where
\begin{eqnarray}\label{eq:eff-bag}
\nonumber    B_{eff}&=&B+\tilde{B}=B+\frac{\epsilon}{1-4\epsilon}(\rho-p_r/c^2-2p_t/c^2)\\
    &=&\left[1+\frac{4\epsilon}{3(1-4\epsilon)}\right]B+\frac{2\epsilon}{3(1-4\epsilon)}(\rho-3p_t/c^2).
\end{eqnarray}
At the GR limit, $\epsilon \to 0$, we obtain $B_{eff}=B$. In our case, for $\epsilon=0.041$, the negativity of the matter energy-stress tensor trace, or the DEC as we call it here, implies the positivity of the effective bag constant $B_{eff}$. At the center ($p_t=p_r$) the effective bag constant reduces to $B_{eff}=\frac{1}{1-4\epsilon}B$, while at the surface ($p_r=0$ and $\rho=\rho_\text{s}=4B$) it gives $B_{eff}=\frac{1}{1-4\epsilon}(B-2\epsilon p_t/c^2)$. The nonvanishing tangential pressure at the surface essentially differentiates RT from GR. Additionally, Equation \eqref{eq:eff-EoS} shows the role of the nonminimal coupling between matter and geometry to effectively mimic bag constant similar to the cases when QCD corrections are included.\\
Now we turn our attention to quark matter physics to check the viability of the model on the microphysics level. In a compact star, the quark chemical potential is $\mu \simeq 500~\text{MeV}$ at most and the temperature is practically zero. Thus, we consider quark matter with three flavors; up, down and strange. For up and down quarks their masses are negligible $m_{u} c^2\simeq m_{d} c^2 \simeq 5~\text{MeV} \ll \mu$, in addition, we neglect the strange quark mass $m_s c^2\simeq 90~\text{MeV}$ for noninteracting case for simplicity, which is still useful as an approximation in the present context. For neutral quark matter with three flavors, the neutrality and the chemical equilibrium constraints render the quark matter symmetric whereas the quark chemical potential $\mu:=\mu_u=\mu_d=\mu_s = \frac{1}{3} \mu_B$ and its density number $n=3 n_B$, in comparison to normal (neutron) baryonic matter, which keeps their product invariant, i.e. $n \mu = n_B \mu_B$. One should assume some model to discuss possible constraints on $\mu$ and $n$, so we use the EoS \eqref{eq:MIT-EoS}, as induced by KB ansatz, where $c_s^2=\frac{1}{3}c^2$ and $\rho_\text{s}=-(c_0/c_1)\rho_\star \approx 2.9\times 10^{14}$~g/cm$^3$ as just derived in the previous section. Clearly the present case represents massless quarks in MIT bag model whereas the bag constant $B=\frac{1}{4}\rho_\text{s}$. The microscopic density of the quarks is given by
\begin{equation}\label{eq:micro-dens}
    \rho c^2= \frac{9}{4\pi^2 (\hbar c)^3}\mu^4+B c^2,
\end{equation}
where $\hbar$ is the reduced Planck's constant. Using Witten stability condition of quark matter, at zero pressure, where the quark density per baryon is less than the energy per baryon of Iron nuclei ($\mu_B=\rho c^2/n_B < 930~\text{MeV}$). Then the stability condition of quark matter, at the surface ($p_r=0$ and $\rho_\text{s}=4B$), finds the maximum quark chemical potential
\begin{equation}
    \mu_{s} < \frac{1}{3} \mu_B \simeq 310~\text{MeV}.
\end{equation}
By virtue of Eq. \eqref{eq:micro-dens} we obtain
\begin{equation}
    \mu_{s}=\left(\frac{\pi^2 \rho_\text{s} c^2}{3}\right)^{1/4} (\hbar c)^{3/4} = \left(\frac{4 \pi^2}{3}\right)^{1/4} (B c^2)^{1/4} (\hbar c)^{3/4} <310~\text{MeV}.
\end{equation}
This puts an upper bound on the bag constant $B^{1/4}< 163\left(c/\hbar\right)^{3/4}~\text{MeV}/c^2$, and
\begin{equation}
    n_{s}=\left(\frac{192}{\pi^2}\right)^{1/4} \left(\frac{B c^2}{c\hbar}\right)^{3/4}.
\end{equation}
On the other hand, in a typical quark star, with EoS as obtained by Equation \eqref{eq:MIT-EoS2}, the surface density is related to the MIT bag constant $\rho_\text{s}=4 B$. In general, the phenomenological parameters $\rho_\text{s}$ and $B$ are considered as free parameters provided that $\rho_\text{s} \geq \rho_\text{nuc}=2.7\times 10^{14}~\text{g/cm}^{3} \simeq 151.46~\text{MeV}/c^2/\text{fm}^3$ , otherwise deconfined quark matter would be visible in atomic nuclei, see \citep{Char:2019wvo}. This sets a lower bound on the bag constant $B^{1/4}>130.9\left(c/\hbar\right)^{3/4}~\text{MeV}/c^2$. Interestingly, in our case, the constants $\rho_\text{s}$ and $B$ are completely determined by the compactness and Rastall parameters $\rho_\text{s}=4 B = -(c_0/c_1) \rho_\star$. In the present model, for the pulsar PSR J0952\textendash{0607}, we have $c_1 \simeq 1/3$, $c_0\simeq -0.36$ and $\rho_\star\simeq 151.83$ MeV/$c^2$/fm$^3$. This gives $B=-\frac{1}{4}(c_0/c_1)\rho_\star$, i.e. $B^{1/4}\simeq 133.57\left(c/\hbar\right)^{3/4}~\text{MeV}/c^2$ in the viable range $130.9\left(c/\hbar\right)^{3/4}~\text{MeV}/c^2 <B^{1/4}< 163\left(c/\hbar\right)^{3/4}~\text{MeV}/c^2$. Notably, by imposing the upper conformal limit constraints on the sound speed within the pulsar PSR J0952\textendash{0607} in GR ($\epsilon \to 0$), we obtained $R=16.41 \pm 1.19$ km which in turn determines the surface density $\rho_\text{s}\approx 118.28$ MeV/$c^2$/fm$^{3}< \rho_\text{nuc}$, i.e. $B^{1/4}=124.3\left(c/\hbar\right)^{3/4}~\text{MeV}/c^2$. This violates the lower bound $B^{1/4}=130.9\left(c/\hbar\right)^{3/4}~\text{MeV}/c^2$ required to avoid visibility of deconfined quark in atomic nuclei. It is obvious that the conformal upper limit on the sound speed implies larger size in GR than in RT, inconsistent with measured radii of relevant star masses as obtain in the previous section, and consequently smaller surface density ($\rho_\text{s}=4B$) violating the lower bound on the bag constant. The present results show the novel role of the matter\textendash{geometry} nonminimal coupling to obtain a stellar model consistent with macroscopic and microscopic physical constraints.

In conclusion, according to our analysis the nonminimal coupling between geometry and matter in Rastall field equations just mimics an effective bag constant of the MIT bag model. Our approach to fix the maximum sound speed at the center to the conjectured conformal bound is compatible with massless quark matter approximation. On the other hand, the observational constraints on the matter-geometry nonminimal coupling term provides numerical values compatible with quark star model with three flavors (up, down and strange), unlike the GR case. In this sense, we find the model at hand successfully relates the macroscopic physics to the microscopic one rendering the model qualitatively comparable, on equal footing, with the popular MIT bag model of strange quark matter.
\section{Crustal Structure and Mass\textendash{Radius} Relation}
\label{Sec:Mass-Radius}

One interesting feature which characterizes NS is the crustal structure. The crust layer occurs where the density goes below neutron drip density ($\simeq 10^{11}$ g/cm$^{3}$ which is $10^{-3}$ below the average density in our case). Since the lowest density at the surface according to the present model is $\sim 2.8\times 10^{14}$ g/cm$^{3}$, there is no chance for the pulsar PSR J0952\textendash{0607} to form a crustal structure. This is also consistent with the formation of quark star as suggested by \cite{Witten:1984rs}, whereas a quark matter will quickly develop the equilibrium strangeness content via weak interactions, such as $u+d \to u+s$. The energy will be lowered as strange quarks are created one by one until equilibrium is reached. Then, the quark matter component can rapidly grow in a neutron star by absorbing free neutrons, since there is no Coulomb barrier within the star. The entire star will turn into quark matter, except for a thin outer crust in which there are no free neutrons \citep{Witten:1984rs}. For an accretion NS or QS, we could expect a crustal structure with density at neutron drip density $\simeq 4\times 10^{11}$ g/cm$^{3}$ at the bottom of the crust, if Coulomb barrier in between is sufficiently large \citep{Zdunik:2001yz}. In this sense, we comfortably use the light and thin crust approximation. Accordingly, we estimate the crust mass and radius for the both static and rotating cases.\\
(1) The crust in nonrotating strange quark stars \citep{Zdunik:2001yz},
\begin{eqnarray}
    \Delta M_\text{cr}^\text{stat}&=&8\pi R^3 \frac{(1-C)}{C}\frac{p_\text{cr}}{c^2},\\
    \Delta R_\text{cr}^\text{stat}&=&\frac{2\gamma}{\gamma-1}\frac{R}{C}\frac{p_\text{cr}}{\rho_\text{cr}c^2},
\end{eqnarray}
where $\rho_\text{cr}$ ($p_\text{cr}$) is the density (pressure) at the bottom of the crust, and $\gamma$ is the average adiabatic index of the outter crust. Following \cite{Zdunik:2001yz}, we assume the solid crust layer is described by BPS model of dense matter below the neutron drip \citep{Baym:ApJ1971}, whereas the density $\rho_\text{cr}=4.3\times 10^{11}$ g/cm$^{3}$ equal to the neutron drip density and the pressure $p_\text{cr}=7.8\times 10^{29}$ dyn/cm$^{2}$ with adiabatic index $\gamma=1.28$. However the adiabatic index for the present anisotropic case is given by \citep{10.1093/mnras/265.3.533}
\begin{equation}\label{eq:adiabatic_index}
    \gamma= \frac{4}{3}\left(1+\frac{F_a}{2 |{dp_r/dr}|}\right)_\text{max},
\end{equation}
where the anisotropic force $F_a:= 2\Delta/r$. Clearly $\gamma = 4/3$ in absence of anisotropy and $\gamma > 4/3$ for strong anisotropic case ($p_t > p_r$). For the BW pulsar PSR J0952\textendash{0607} we use the field equations \eqref{eq:Feqs2} and Equation \eqref{eq:adiabatic_index} to evaluate the adiabatic index at the surface $\gamma\approx 1.7$. This gives a crustal mass $\Delta M_\text{cr}^\text{stat}\approx 3.14\times 10^{-5}\, M_\odot$ and thickness $\Delta R_\text{cr}^\text{stat}\approx 0.28$ km.\\
(2) The crust in rotating strange quark stars \citep{Zdunik:2001yz},
\begin{eqnarray}
    \Delta M_\text{cr}^\text{rot}&=&\Delta M_\text{cr}^\text{stat}\left(1+0.24 \left(f/f_\text{Kep}\right)^2 + 0.16 \left(f/f_\text{Kep}\right)^8\right),\qquad \\
    \Delta R_\text{cr(eq)}^\text{rot}&=&\Delta R_\text{cr}^\text{stat}\left(1+0.4 \left(f/f_\text{Kep}\right)^2 + 0.3 \left(f/f_\text{Kep}\right)^6\right).
\end{eqnarray}
where the Keplerian frequency in the above formulae $f_\text{Kep}=1$ kHz. This gives $\Delta M_\text{cr}^\text{rot} \approx 3.54\times 10^{-5}\, M_\odot$ and equatorial thickness $\Delta R_\text{cr(eq)}^\text{rot}\approx 0.34$ km. As expected for light and thin crust the change in the mass is negligible and in the radius $\Delta R\lesssim 0.34$ km much smaller than the uncertainty of our estimated value $R \approx 14.1 \pm 1$ km. In the following, we are going to confront the mass\textendash{radius} relation corresponds to the present model with several observational measurements.

\begin{figure}[ht!]
\includegraphics[width=0.45\textwidth]{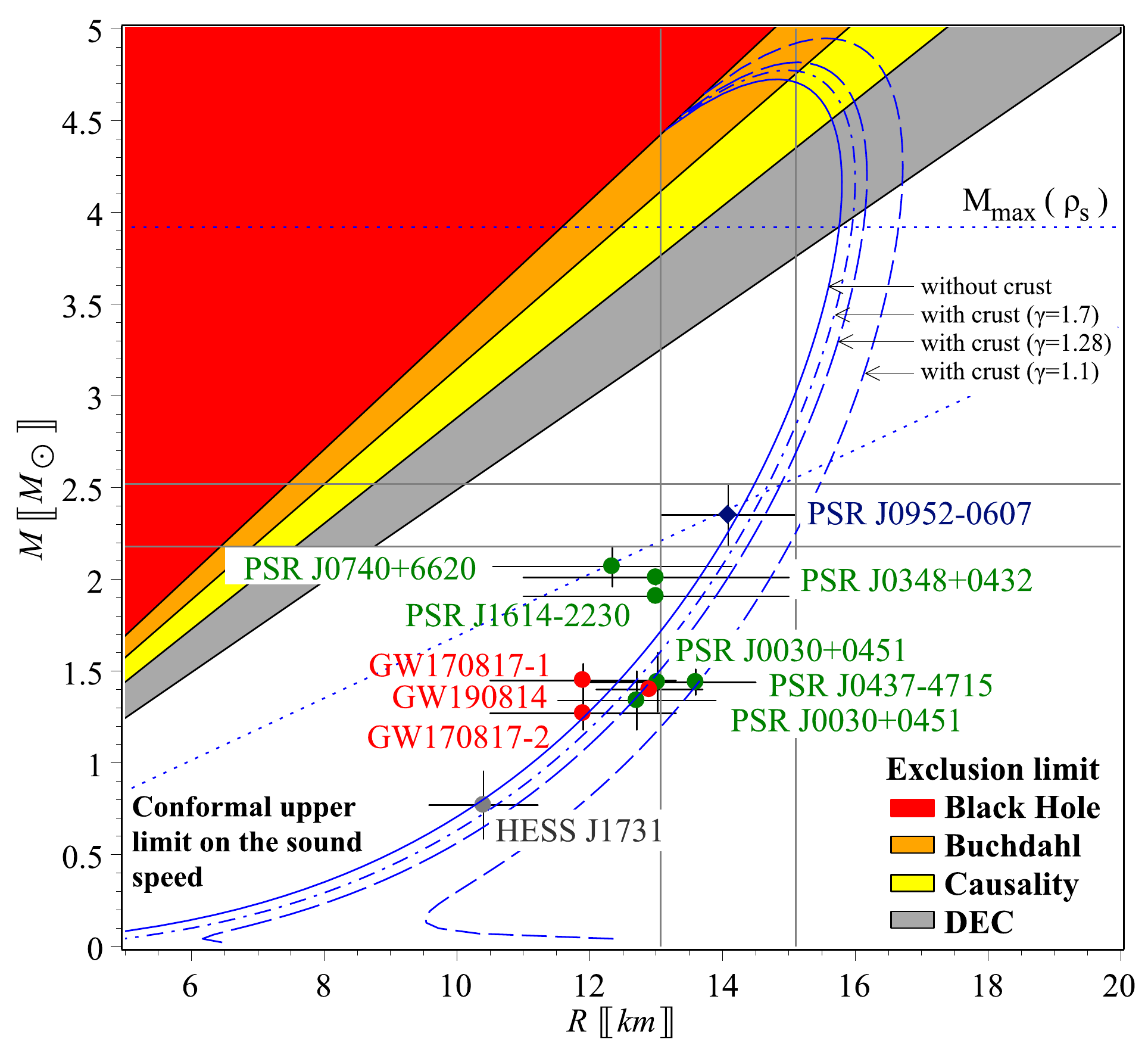}
\caption{The mass\textendash{radius} diagram as predicted by the present model for a density $\rho_\text{s}=2.8\times 10^{14}$ g/cm$^3$ at the boundary of a compact star with no crust as given by the solid curve. The crust contribution is given for different values of adiabatic indices. The horizontal dotted line represents the upper mass\textendash{radius} values as allowed by the DEC, $\rho c^2-p_r-2p_t \geq 0$, which determines $M_\text{max}=3.92 M_\odot$ (blue\textendash{dotted} horizontal line) at a maximum radius of $R_\text{max}=15.76$ km. Exclusion limits due to Table \ref{tab:comp-limits} are given in different colors as indicated by the legend. The conformal sound speed upper limit constraint $c_s^2/c^2 \lesssim 1/3$ is represented by the dot diagonal line. The Navy solid diamond represents the mean values of the mass\textendash{radius} of the heaviest pulsar PSR J0952\textendash{0607}, the red solid circles represent LIGO and Virgo constraints on the mass\textendash{radius} values, the green solid circles represent NICER constraints and the gray solid circles represents the lightest pulsar at the center of the supernova remnant HESS J1731\textendash{347}. Clearly all pulsars satisfy the conformal sound speed limit and perfectly fit with the boundary density condition.
\label{Fig:MR-Reln}}
\end{figure}

We take the boundary condition $\rho_\text{s}=2.8\times 10^{14}$ g/cm$^3$ at the surface of the compact object, for arbitrary values of the compactness parameter $0 \leq C \leq 1$, and by solving the density profile, Equation \eqref{eq:Feqs2}, for the radius $R$, we calculate the corresponding gravitational mass as determined by the matching conditions \eqref{eq:bo}, i.e. $M= \frac{c^2 R}{2 G} (1-e^{-a_2})$. Thus, we give the mass\textendash{radius} curve for the boundary density condition $\rho=\rho_\text{s}$ as seen by the mass\textendash{radius} diagram in Figure \ref{Fig:MR-Reln}. Clearly the curve is arbitrarily extendable to a maximum compactness approaching the black hole limit $C\to 1$ where the red region represents a region at which this limit is violated. In fact this is common in anisotropic fluid models even in GR, c.f. \citep{Alho:2022bki}. Keeping in mind the effective (total) matter in RT is having the same characteristics of the matter sector in GR as obtained by Equation \eqref{RT2}. Thus the result would not be altered in RT gravity, since black hole limit is the same in both theories. However for other physical constraints different bounds on the compactness value are obtained in Table \ref{tab:comp-limits}.\\
\begin{table}
\caption{Physical bounds on the maximum compactness in Rastall gravity ($\epsilon=0.041$) in comparison to GR ($\epsilon=0$).}
    \label{tab:comp-limits}
   \begin{center}
\begin{tabular}{lcc}
\hline
 limit  & GR ($\epsilon=0)$ & RT ($\epsilon=0.041$) \\
        & $C_\text{max}$  & $C_\text{max}$\\
\hline
 BH limit & 1 & 1 \\
 Buchdahl limit (fluid)        & $8/9 \simeq 0.89$ & 0.93  \\
 Causality (maximally compact) & 0.851             & 0.857 \\
 DEC                           & 0.715             & 0.735 \\
 Conformal limit               & 0.423             & 0.493 \\
 \hline
\end{tabular}
\end{center}
\end{table}
We briefly discuss the different limits given in Table \ref{tab:comp-limits}. For the well known general relativistic Buchdal limit, $C\leq 8/9$ (i.e. $M/R\leq \frac{4 c^2}{9 G}$) as obtained by \citep{PhysRev.116.1027}, this limit is specifically derived for isotropic fluid ($p_r= p_t$) or mildly anisotropic ($p_r\gtrsim p_t$) with spherically symmetric spacetime configuration. Therefore, Buchdahl’s limit can be easily violated by relaxing one or more of these assumptions. In our case, it has been shown that the fluid maximum compactness reaches $C^\text{fluid}_\text{max}\approx 0.93$ in RT above GR Buchdahl limit \citep{ElHanafy:2022kjl}. The orange region in Figure \ref{Fig:MR-Reln} represents a region at which this limit is violated. Similar to our case, by relaxing mild anisotropy condition, Buchdahl limit is violated for strong anisotropic stars by introducing elasticity for instance, $C\approx 0.924$,  even in GR \citep{Alho:2022bki,Alho:2021sli}, see also \citep{Raposo:2018rjn,Cardoso:2019rvt}. Also, we impose the causality constraint by accounting for the sound speed limit $c_s=c$ within a maximally compact object with EoS $p_r(\rho)=c_s^2(\rho-\rho_\text{s})$. We find a maximum compactness $C=0.851 (0.857)$ in the GR (RT) case. The yellow region in Figure \ref{Fig:MR-Reln} represents a region at which this limit is violated. Finally, we include the maximum $C$ values by accounting for the DEC, $\rho c^2-p_r-2p_t \geq 0$, which obtains the GR value $C= 0.715$ as previously obtained \citep{Roupas:2020mvs} for KB model, while it gives $C= 0.735$ in RT with $\epsilon=0.041$ \citep{ElHanafy:2022kjl}. Remarkably, the obtained values by imposing the DEC are close to those of Buchdahl limit, for mild anisotropic fluid, when strengthen by causality \citep{Alho:2021sli,Boskovic:2021nfs,Urbano:2018nrs}. The dotted diagonal line represents the mass\textendash{radius} limit corresponds to the conformal upper limit on the sound speed $c_s=c/\sqrt{3}$, whilst the horizontal dashed line sets the upper limit on the mass $M(\rho_\text{s})=3.92 M_\odot$ as allowed by the DEC for the boundary density $\rho_\text{s}$.

In fact the stellar evolution of isotropic spheres, in general, leaves the so-called lower mass gap between heaviest NS and lightest BH, $2.2 M_\odot \lesssim M \lesssim 5 M_\odot$, unpopulated \citep[c.f.][]{Yang:2020xyi}. The ability of the present model to produce compact objects in this range is due to strong anisotropies as assumed for the stress-energy tensor, Equation \eqref{Tmn-anisotropy}. We note that anisotropies in matter fields arise naturally at high densities due to superfluidity, solidification, strong magnetic fields, hyperons, quarks as well as pion and kaon condensation. Therefore, presence of anisotropic pressures is more realistic than the ideally isotropic case. Strong anistotropies induce additional repulsive force allowing stars to gain more compactness. In this sense, anisotropic ultracompact star model is not the exceptional case. The ability of anisotropic fluid models to produce compactness $C\to 1$ is suppressed by the exclusion limits due to different physical constraints. The DEC represents the most restrictive condition which is, in fact, related to the negativity of the trace of stress-energy tensor.

The mass\textendash{radius} curve, for the boundary density $\rho_\text{s}$, in Figure \ref{Fig:MR-Reln} shows a perfect agreement with our estimated value of the PSR J0952\textendash{0607} radius $R=14.087 \pm 1.0186$ km corresponds to the observed mass $M=2.35 \pm 0.17 M_\odot$. The solid curve indicates the case when no crust is assumed while other curves include the crust contribution for different values of adiabatic indices. Noting that the relevant value in our case is given by the $\gamma=1.7$ curve with slight change towards higher radii. For adiabatic indices lower than $\gamma<4/3$, i.e. Newtonian approximation or mildly anisotropic star, the crust thickness becomes larger with extreme change at low masses $M\lesssim 0.2 M_\odot$.

We also include additional measurements of mass\textendash{radius} of several MSPs with mass $\simeq 2 M_\odot$: The PSR J0740+6620 with mass $M= 2.07 \pm 0.11 M_\odot$ and radius $R=12.34^{+1.89}_{-1.67}$ km as inferred by Gaussian process of a nonparameteric EoS approach using NICER+XMM data \citep{Legred:2021hdx}, the PSR J0348+0432 with mass $M= 2.01 \pm 0.04 M_\odot$ and an estimated radius $R=13 \pm 2$ km \citep{Antoniadis:2013pzd} and the PSR J1614\textendash{2230} with mass $M= 1.908 \pm 0.016 M_\odot$ and radius $R=13 \pm 2$ km \citep{Demorest:2010bx,Fonseca:2016tux,NANOGRAV:2018hou} using the relativistic Shapiro time delay technique. Also, we include more mass\textendash{radius} measurements of MSPs with relatively small masses $\simeq 1.5 M_\odot$: The PSR J0030+0451 with mass $M=1.44^{+0.15}_{-0.14} M_\odot$ and radius $R= 13.02_{-1.06}^{+1.24}$ km as measured by NICER \citep{Miller:2019cac}, with another independent NICER measurement with mass $M= 1.34^{+0.15}_{-0.16} M_\odot$ and radius $R= 12.71^{+1.14}_{-1.19}$ km \citep{Raaijmakers:2019qny}. The PSR J0437\textendash{4715} with mass $M=1.44 \pm 0.07 M_\odot$ \citep{Reardon:2015kba} and radius $R=13.6 \pm 0.9$ km \citep{Gonzalez-Caniulef:2019wzi} by the analyses of the surface X-ray thermal emission. Moreover, we include three observed mass\textendash{radius} values as inferred by gravitational wave signals as detected by LIGO/Virgo collaboration: The first detected NS-NS merger GW170817-1 with mass $M=1.45 \pm 0.09 M_\odot$ and radius $R=11.9 \pm 1.4$ km, and GW170817-2 with mass $M=1.27 \pm 0.09 M_\odot$ and radius $R=11.9 \pm 1.4$ km \citep{LIGOScientific:2018cki}. The LIGO/Virgo constraints on the radius of a canonical NS using GW170817+GW190814 signals with mass $M=1.4 M_\odot$ and radius $R=12.9\pm 0.8$ km \citep{LIGOScientific:2020zkf}.

Additionally, we include the mass and radius measurements of the possibly lightest NS, $M=0.77_{-0.17}^{+0.20}~M_{\odot}$ and $R=10.4^{+0.86}_{-0.78}$ km, located within the supernova remnant HESS J1731\textendash{347} based on the X-ray spectrum and distance estimate as obtained by Gaia observations \citep{2022NatAs...6.1444D}. Indeed for light compact objects with hadron matter EoS, as mass decreases, the radius in fact increases much larger than what is already observed $R\approx 10.4$ km, whereas the extension of the mass\textendash{radius} towards large values of radii violates the realistic mass-shedding limit for the highest known pulsar frequency, 716 Hz, for PSR J1748\textendash{2446ad}. On the contrary, in our case, in agreement with strange quark star, as mass decreases, the radius decreases too. The present model fits perfectly this light compact object which may strengthen the suggestion of this object being a strange quark star, whereas the mass\textendash{radius} curve corresponds to strange quark matter is extremely different from hadrons one at these low mass limits.

In conclusion, the curve perfectly fits wide range of well observed mass\textendash{radius} values as obtained by NICER and LIGO/Virgo observations in agreement with our predicted radius $R=14.087 \pm 1.0186$ km for the heaviest pulsar PSR J0952\textendash{0607} in addition to lightest compact objects such as the supernova remnant HESS J1731\textendash{347}. This confirms that the same physics can be applied for those compact stars while keeping the sound speed below its conformal limit everywhere inside the star. Remarkably, our results could interpret the companion of the merger GW190814, with $M=2.6 M_\odot$ \citep{LIGOScientific:2020zkf}, in the lower mass gap, as anisotropic compact star.

\section{Conclusion}\label{Sec:Conclusion}
Recent observations by NICER show that the pulsar PSR J0740+6020 has much more mass than PSR J0030+0451 but both are having almost same size $R\simeq 13$ km. In addition, existence of compact stars with masses $M\gtrsim 2 M_\odot$ is in favor of violation of the sound speed upper limit $c^2_s=c^2/3$. For a non-rotating NS model, we argue that (i) High dense matter, at the core of a massive NS, is expected to generate anisotropic pressures. This induces a repulsive anisotropic force as indicated by TOV equation where $p_t > p_r$, which in turn supports the massive NS not to collapse under gravitation attraction to smaller size. (ii) The conjectured conformal sound speed bound can be applied everywhere inside an NS within a nonminimal matter\textendash{geometry} coupling framework as in Rastall gravity unlike GR.

We take Rastall parameter $\epsilon=0.041$, which quantifies the matter\textendash{geometry} nonminimal coupling, as inferred by NICER+XMM observations of the pulsar PSR J0740+6020. For the BW pulsar PSR J0952\textendash{0607} with an observed mass $M=2.35 \pm 0.17 M_\odot$, we assume that the sound speed upper limit is applied at the center where $c_s^2$ is maximized. We investigate the nonmimimal matter\textendash{geometry} coupling in Rastall gravity showing its role to effectively mimic the bag constant of the the popular MIT bag model. In this sense, we investigate the model on the microscopic level in comparison to strange quark star models. In this study we show that the KB model in general relates the pressure and density inducing a linear EoS, $p(\rho)=c_s^2 (\rho - \rho_\text{s})$, which renders the model qualitatively comparable, on equal footing, with the popular MIT bag model of quark matter whereas $\rho_\text{s}=4B$. We estimate its radius $R=14.087 \pm 1.0186$ km in agreement with other observations of $\sim 2 M_\odot$ pulsars which measure $R \sim 13-14$ km. The model determines a surface density $\rho_\text{s}=2.8\times 10^{14}$ g/cm$^3$ above the the nuclear saturation density, which consequently determines a viable bag constant value $B^{1/4} \approx 134\left(c/\hbar\right)^{3/4}~\text{MeV}/c^2$ consistent with the lower bound $B^{1/4}=130.9\left(c/\hbar\right)^{3/4}~\text{MeV}/c^2$ required to avoid visibility of deconfined quark in atomic nuclei. Notably, the GR treatment, $\epsilon=0$, following same procedure obtains the BW size $R=16.41 \pm 1.19$ km much larger than the measured radii of the comparable pulsars' masses. This determines a surface density $\rho_\text{s}\approx 118$ g/cm$^3$ less than the saturation nuclear density and consequently $B^{1/4}\approx 124\left(c/\hbar\right)^{3/4}~\text{MeV}/c^2$ which violates the lower bound constraint. On the other hand, if $R\sim 14$ km is hold in GR the sound speed necessarily exceeds the conjectured conformal upper limit. The present study may point out some difficulties of the GR gravity to keep the conformal limit of the sound speed valid. It, additionally, shows the novel role of the matter\textendash{geometry} nonminimal coupling to obtain a stellar model consistent with macroscopic and microscopic physical constraints.

We obtain the corresponding mass\textendash{radius} diagram, which has been shown to be in a perfect agreement with a wide range of astrophysical observations from lightest to heaviest pulsars observed. Those includes our estimated size of the BW pulsar PSR J0952\textendash{0607} and other observations of several pulsars as obtained by NICER and LIGO/Virgo observations. Remarkably the curve fits the light compact object of the supernova remnant HESS J1731\textendash{347} whereas radius decreases as mass decreases, in favor of quark star model, unlike normal matter with hadron EoS. This ensures that same physics can be applied for those pulsars, regardless their masses, while keeping the sound speed below its conformal limit everywhere inside the stars.

\appendix

\section{Impact of static approximation on the radius measurement}
\label{App:Radius_static_approx}

In Section \ref{Sec:Model}, using the quasi-universal relation \eqref{eq:BR_formula} for the pulsar PSR J0952\textendash{0607}, we showed that the mass increase due to rotational frequency is $\simeq 3.5\%$ much less than the mass uncertainty, $M=2.35 \pm 0.17 M_\odot$, which justifies the static approximation \eqref{eq:metric}. In Section \ref{Sec:Radius}, we estimated the pulsar radius $R=14.087 \pm 1.0186$ km (with a relative error $\Delta R/R \simeq 7.23\%$) as inferred by Rastall gravity assuming that the speed of sound is consistent with conformal upper limit as induced by the perturbative QCD at high densities. Although the error in the radius is induced by the error in the mass measurement in the present case, we discuss the impact of the static approximation on the radius in general.

As is shown no significant change is expected in mass measurements due to rotational frequency, similar conclusion has been derived by NICER for the radius as well in the case of the pulsar PSR J0030+0451 \citep[see][Fig. 13]{Miller:2019cac}. In particular, the M\textendash{R} comparison between nonrotating spherical symmetry case and rotating case (with $f=205$ Hz) for several EoSs shows no significant change in the radius too specially for heavy compact stars with $M> 2M_\odot$. Similar conclusion has been derived for realistic NS parameters ($1.4M_\odot$, $12$ km, $600$ Hz). It has been shown that the impact of the rotating and nonrotating cases on the radius is on the order of $4\%$ \citep{Baubock:2014xha}. Assuming that the least square quadratic fit \cite[Eq. (27)]{Baubock:2014xha},
\begin{eqnarray}
\nonumber    \frac{\Delta R}{R}&=& \left[\left(0.108-0.096\frac{M}{M_\odot}\right)+\left(-0.061+0.114\frac{M}{M_\odot}\right)\frac{R}{10~\text{km}}\right.\\
\nonumber    &&-0.128 \left.\left(\frac{R}{10~\text{km}}\right)^2\right]\left(\frac{f}{1000~\text{Hz}}\right)^2,
\end{eqnarray}
is applied in the present study for the BW pulsar PSR J0952\textendash{0607}, we get a similar result $\Delta R/R \lesssim 4\%$.

Furthermore, one may consider errors in the measuring value of the moment of inertia $\Delta I/I$ and its impact on the radius error as in the following formula \cite[see][appendix A]{Breu:2016ufb}
\begin{equation}
\nonumber    \frac{\Delta R}{R}= f(\mathcal{C}) \frac{\Delta I}{I},
\end{equation}
where
\begin{equation}
\nonumber    f(\mathcal C)=\frac{1+(\Tilde{a}_1/\Tilde{a}_0)\mathcal{C}+(\Tilde{a}_4/\Tilde{a}_0)\mathcal{C}^4}{2+(\Tilde{a}_1/\Tilde{a}_0)\mathcal{C}-2(\Tilde{a}_4/\Tilde{a}_0)\mathcal{C}^4}.
\end{equation}
Given that $\Tilde{a}_0=0.244$, $\Tilde{a}_1=0.638$, $\Tilde{a}_4=3.202$ \citep{Breu:2016ufb} and $\mathcal{C}:=C/2=\frac{GM}{c^2R}\simeq 0.248$ as estimated in the present study, and by taking the relative error in the moment of inertia $\simeq 10\%$ \citep[see][Fig. 2]{Breu:2016ufb}, we estimate $\Delta R/R \simeq 6.5\%$.

Finally, we find that the above calculations are very useful to prove the static assumption justified/unjustified when the radius measurement is provided.



\begin{thebibliography}{}
\expandafter\ifx\csname natexlab\endcsname\relax\def\natexlab#1{#1}\fi
\providecommand{\url}[1]{\href{#1}{#1}}
\providecommand{\dodoi}[1]{doi:~\href{http://doi.org/#1}{\nolinkurl{#1}}}
\providecommand{\doeprint}[1]{\href{http://ascl.net/#1}{\nolinkurl{http://ascl.net/#1}}}
\providecommand{\doarXiv}[1]{\href{https://arxiv.org/abs/#1}{\nolinkurl{https://arxiv.org/abs/#1}}}

\bibitem[{Abbott {et~al.}(2017)}]{TheLIGOScientific:2017qsa}
Abbott, B.~P., {et~al.} 2017, Phys. Rev. Lett., 119, 161101,
  \dodoi{10.1103/PhysRevLett.119.161101}

\bibitem[{Abbott {et~al.}(2018)}]{LIGOScientific:2018cki}
---. 2018, Phys. Rev. Lett., 121, 161101,
  \dodoi{10.1103/PhysRevLett.121.161101}

\bibitem[{Abbott {et~al.}(2020{\natexlab{a}})}]{LIGOScientific:2020aai}
---. 2020{\natexlab{a}}, Astrophys. J. Lett., 892,
  \dodoi{10.3847/2041-8213/ab75f5}

\bibitem[{Abbott {et~al.}(2020{\natexlab{b}})}]{LIGOScientific:2020zkf}
Abbott, R., {et~al.} 2020{\natexlab{b}}, Astrophys. J. Lett., 896,
  \dodoi{10.3847/2041-8213/ab960f}

\bibitem[{Alho {et~al.}(2022{\natexlab{a}})Alho, Nat\'ario, Pani, \&
  Raposo}]{Alho:2022bki}
Alho, A., Nat\'ario, J., Pani, P., \& Raposo, G. 2022{\natexlab{a}}, Phys. Rev.
  D, 106, \dodoi{10.1103/PhysRevD.106.L041502}

\bibitem[{Alho {et~al.}(2022{\natexlab{b}})Alho, Nat\'ario, Pani, \&
  Raposo}]{Alho:2021sli}
---. 2022{\natexlab{b}}, Phys. Rev. D, 105, 044025,
  \dodoi{10.1103/PhysRevD.105.044025}

\bibitem[{Alsing {et~al.}(2018)Alsing, Silva, \& Berti}]{Alsing:2017bbc}
Alsing, J., Silva, H.~O., \& Berti, E. 2018, Mon. Not. Roy. Astron. Soc., 478,
  1377, \dodoi{10.1093/mnras/sty1065}

\bibitem[{Altiparmak {et~al.}(2022)Altiparmak, Ecker, \&
  Rezzolla}]{Altiparmak:2022bke}
Altiparmak, S., Ecker, C., \& Rezzolla, L. 2022, Astrophys. J. Lett., 939,
  \dodoi{10.3847/2041-8213/ac9b2a}

\bibitem[{Annala {et~al.}(2020)Annala, Gorda, Kurkela, N\"attil\"a, \&
  Vuorinen}]{Annala:2019puf}
Annala, E., Gorda, T., Kurkela, A., N\"attil\"a, J., \& Vuorinen, A. 2020,
  Nature Phys., 16, 907, \dodoi{10.1038/s41567-020-0914-9}

\bibitem[{Antoniadis {et~al.}(2013)}]{Antoniadis:2013pzd}
Antoniadis, J., {et~al.} 2013, Science, 340, 6131,
  \dodoi{10.1126/science.1233232}

\bibitem[{Arzoumanian {et~al.}(2018)}]{NANOGRAV:2018hou}
Arzoumanian, Z., {et~al.} 2018, Astrophys. J., 859, 47,
  \dodoi{10.3847/1538-4357/aabd3b}

\bibitem[{Baubock {et~al.}(2015)Baubock, Ozel, Psaltis, \&
  Morsink}]{Baubock:2014xha}
Baubock, M., Ozel, F., Psaltis, D., \& Morsink, S.~M. 2015, Astrophys. J., 799,
  22, \dodoi{10.1088/0004-637X/799/1/22}

\bibitem[{{Baym} {et~al.}(1971){Baym}, {Pethick}, \&
  {Sutherland}}]{Baym:ApJ1971}
{Baym}, G., {Pethick}, C., \& {Sutherland}, P. 1971, Astrophys. J., 170, 299,
  \dodoi{10.1086/151216}

\bibitem[{Bedaque \& Steiner(2015)}]{Bedaque:2014sqa}
Bedaque, P., \& Steiner, A.~W. 2015, Phys. Rev. Lett., 114, 031103,
  \dodoi{10.1103/PhysRevLett.114.031103}

\bibitem[{Bhattacharyya {et~al.}(2016)Bhattacharyya, Bombaci, Logoteta, \&
  Thampan}]{Bhattacharyya:2016kte}
Bhattacharyya, S., Bombaci, I., Logoteta, D., \& Thampan, A.~V. 2016, Mon. Not.
  Roy. Astron. Soc., 457, 3101, \dodoi{10.1093/mnras/stw206}

\bibitem[{Bo\v{s}kovi\'c \& Barausse(2022)}]{Boskovic:2021nfs}
Bo\v{s}kovi\'c, M., \& Barausse, E. 2022, JCAP, 02, 032,
  \dodoi{10.1088/1475-7516/2022/02/032}

\bibitem[{Bowers \& Liang(1974)}]{bowers1974anisotropic}
Bowers, R.~L., \& Liang, E. 1974, The Astrophysical Journal, 188, 657

\bibitem[{Bozzola {et~al.}(2019)Bozzola, Espino, Lewin, \&
  Paschalidis}]{Bozzola:2019tit}
Bozzola, G., Espino, P.~L., Lewin, C.~D., \& Paschalidis, V. 2019, Eur. Phys.
  J. A, 55, 149, \dodoi{10.1140/epja/i2019-12831-2}

\bibitem[{Breu \& Rezzolla(2016)}]{Breu:2016ufb}
Breu, C., \& Rezzolla, L. 2016, Mon. Not. Roy. Astron. Soc., 459, 646,
  \dodoi{10.1093/mnras/stw575}

\bibitem[{Buchdahl(1959)}]{PhysRev.116.1027}
Buchdahl, H.~A. 1959, Phys. Rev., 116, 1027, \dodoi{10.1103/PhysRev.116.1027}

\bibitem[{Cardoso \& Pani(2019)}]{Cardoso:2019rvt}
Cardoso, V., \& Pani, P. 2019, Living Rev. Rel., 22, 4,
  \dodoi{10.1007/s41114-019-0020-4}

\bibitem[{Chan {et~al.}(1993)Chan, Herrera, \&
  Santos}]{10.1093/mnras/265.3.533}
Chan, R., Herrera, L., \& Santos, N.~O. 1993, Monthly Notices of the Royal
  Astronomical Society, 265, 533, \dodoi{10.1093/mnras/265.3.533}

\bibitem[{Char {et~al.}(2019)Char, Drago, \& Pagliara}]{Char:2019wvo}
Char, P., Drago, A., \& Pagliara, G. 2019, AIP Conf. Proc., 2127, 020026,
  \dodoi{10.1063/1.5117816}

\bibitem[{Cherman {et~al.}(2009)Cherman, Cohen, \& Nellore}]{Cherman:2009tw}
Cherman, A., Cohen, T.~D., \& Nellore, A. 2009, Phys. Rev. D, 80, 066003,
  \dodoi{10.1103/PhysRevD.80.066003}

\bibitem[{Cromartie {et~al.}(2019)}]{NANOGrav:2019jur}
Cromartie, H.~T., {et~al.} 2019, Nature Astron., 4, 72,
  \dodoi{10.1038/s41550-019-0880-2}

\bibitem[{Darabi {et~al.}(2018)Darabi, Moradpour, Licata, Heydarzade, \&
  Corda}]{Darabi:2017coc}
Darabi, F., Moradpour, H., Licata, I., Heydarzade, Y., \& Corda, C. 2018, Eur.
  Phys. J. C, 78, 25, \dodoi{10.1140/epjc/s10052-017-5502-5}

\bibitem[{De~Pietri {et~al.}(2019)De~Pietri, Drago, Feo, Pagliara, Pasquali,
  Traversi, \& Wiktorowicz}]{DePietri:2019khb}
De~Pietri, R., Drago, A., Feo, A., {et~al.} 2019, Astrophys. J., 881, 122,
  \dodoi{10.3847/1538-4357/ab2fd0}

\bibitem[{Demircik {et~al.}(2021)Demircik, Ecker, \&
  J\"arvinen}]{Demircik:2020jkc}
Demircik, T., Ecker, C., \& J\"arvinen, M. 2021, Astrophys. J. Lett., 907,
  \dodoi{10.3847/2041-8213/abd853}

\bibitem[{Demorest {et~al.}(2010)Demorest, Pennucci, Ransom, Roberts, \&
  Hessels}]{Demorest:2010bx}
Demorest, P., Pennucci, T., Ransom, S., Roberts, M., \& Hessels, J. 2010,
  Nature, 467, 1081, \dodoi{10.1038/nature09466}

\bibitem[{{Doroshenko} {et~al.}(2022){Doroshenko}, {Suleimanov},
  {P{\"u}hlhofer}, \& {Santangelo}}]{2022NatAs...6.1444D}
{Doroshenko}, V., {Suleimanov}, V., {P{\"u}hlhofer}, G., \& {Santangelo}, A.
  2022, Nature Astronomy, 6, 1444, \dodoi{10.1038/s41550-022-01800-1}

\bibitem[{Drago {et~al.}(2014)Drago, Lavagno, \& Pagliara}]{Drago:2013fsa}
Drago, A., Lavagno, A., \& Pagliara, G. 2014, Phys. Rev. D, 89, 043014,
  \dodoi{10.1103/PhysRevD.89.043014}

\bibitem[{Drago {et~al.}(2016)Drago, Lavagno, Pagliara, \&
  Pigato}]{Drago:2015cea}
Drago, A., Lavagno, A., Pagliara, G., \& Pigato, D. 2016, Eur. Phys. J. A, 52,
  40, \dodoi{10.1140/epja/i2016-16040-3}

\bibitem[{Drischler {et~al.}(2021)Drischler, Han, Lattimer, Prakash, Reddy, \&
  Zhao}]{Drischler:2020fvz}
Drischler, C., Han, S., Lattimer, J.~M., {et~al.} 2021, Phys. Rev. C, 103,
  045808, \dodoi{10.1103/PhysRevC.103.045808}

\bibitem[{Drischler {et~al.}(2022)Drischler, Han, \& Reddy}]{Drischler:2021bup}
Drischler, C., Han, S., \& Reddy, S. 2022, Phys. Rev. C, 105, 035808,
  \dodoi{10.1103/PhysRevC.105.035808}

\bibitem[{Ecker \& Rezzolla(2022)}]{Ecker:2022dlg}
Ecker, C., \& Rezzolla, L. 2022, Mon. Not. Roy. Astron. Soc., 519, 2615,
  \dodoi{10.1093/mnras/stac3755}

\bibitem[{El~Hanafy(2022)}]{ElHanafy:2022kjl}
El~Hanafy, W. 2022, Astrophys. J., 940, 51, \dodoi{10.3847/1538-4357/ac9410}

\bibitem[{Farhi \& Jaffe(1984)}]{Farhi:1984qu}
Farhi, E., \& Jaffe, R.~L. 1984, Phys. Rev. D, 30, 2379,
  \dodoi{10.1103/PhysRevD.30.2379}

\bibitem[{Fonseca {et~al.}(2016)}]{Fonseca:2016tux}
Fonseca, E., {et~al.} 2016, Astrophys. J., 832, 167,
  \dodoi{10.3847/0004-637X/832/2/167}

\bibitem[{Fonseca {et~al.}(2021)}]{Fonseca:2021wxt}
---. 2021, Astrophys. J. Lett., 915, \dodoi{10.3847/2041-8213/ac03b8}

\bibitem[{Gonzalez-Caniulef {et~al.}(2019)Gonzalez-Caniulef, Guillot, \&
  Reisenegger}]{Gonzalez-Caniulef:2019wzi}
Gonzalez-Caniulef, D., Guillot, S., \& Reisenegger, A. 2019, Mon. Not. Roy.
  Astron. Soc., 490, 5848, \dodoi{10.1093/mnras/stz2941}

\bibitem[{Herrera \& Santos(1997)}]{herrera1997local}
Herrera, L., \& Santos, N.~O. 1997, Physics Reports, 286, 53

\bibitem[{Krori \& Barua(1975)}]{Krori1975ASS}
Krori, K.~D., \& Barua, J. 1975, Journal of Physics A, 8, 508

\bibitem[{Landry {et~al.}(2020)Landry, Essick, \&
  Chatziioannou}]{Landry:2020vaw}
Landry, P., Essick, R., \& Chatziioannou, K. 2020, Phys. Rev. D, 101, 123007,
  \dodoi{10.1103/PhysRevD.101.123007}

\bibitem[{Legred {et~al.}(2021)Legred, Chatziioannou, Essick, Han, \&
  Landry}]{Legred:2021hdx}
Legred, I., Chatziioannou, K., Essick, R., Han, S., \& Landry, P. 2021, Phys.
  Rev. D, 104, 063003, \dodoi{10.1103/PhysRevD.104.063003}

\bibitem[{Li {et~al.}(2019)Li, Wang, Xu, \& Guo}]{Li:2019jkv}
Li, R., Wang, J., Xu, Z., \& Guo, X. 2019, Mon. Not. Roy. Astron. Soc., 486,
  2407, \dodoi{10.1093/mnras/stz967}

\bibitem[{Lin \& Qian(2020)}]{Lin:2020fue}
Lin, K., \& Qian, W.-L. 2020, Eur. Phys. J. C, 80, 561,
  \dodoi{10.1140/epjc/s10052-020-8116-2}

\bibitem[{McLerran \& Reddy(2019)}]{McLerran:2018hbz}
McLerran, L., \& Reddy, S. 2019, Phys. Rev. Lett., 122, 122701,
  \dodoi{10.1103/PhysRevLett.122.122701}

\bibitem[{Miller {et~al.}(2019)}]{Miller:2019cac}
Miller, M.~C., {et~al.} 2019, Astrophys. J. Lett., 887,
  \dodoi{10.3847/2041-8213/ab50c5}

\bibitem[{Miller {et~al.}(2021)}]{Miller:2021qha}
---. 2021, Astrophys. J. Lett., 918, \dodoi{10.3847/2041-8213/ac089b}

\bibitem[{Moradpour {et~al.}(2017{\natexlab{a}})Moradpour, Bonilla, Abreu, \&
  Neto}]{Moradpour:2017ycq}
Moradpour, H., Bonilla, A., Abreu, E. M.~C., \& Neto, J.~A. 2017{\natexlab{a}},
  Phys. Rev. D, 96, 123504, \dodoi{10.1103/PhysRevD.96.123504}

\bibitem[{Moradpour {et~al.}(2017{\natexlab{b}})Moradpour, Sadeghnezhad, \&
  Hendi}]{Moradpour:2016ubd}
Moradpour, H., Sadeghnezhad, N., \& Hendi, S.~H. 2017{\natexlab{b}}, Can. J.
  Phys., 95, 1257, \dodoi{10.1139/cjp-2017-0040}

\bibitem[{Moustakidis {et~al.}(2017)Moustakidis, Gaitanos, Margaritis, \&
  Lalazissis}]{Moustakidis:2016sab}
Moustakidis, C.~C., Gaitanos, T., Margaritis, C., \& Lalazissis, G.~A. 2017,
  Phys. Rev. C, 95, 045801, \dodoi{10.1103/PhysRevC.95.045801}

\bibitem[{Nashed \& El~Hanafy(2022)}]{Nashed:2022zyi}
Nashed, G. G.~L., \& El~Hanafy, W. 2022, Eur. Phys. J. C, 82, 679,
  \dodoi{10.1140/epjc/s10052-022-10634-0}

\bibitem[{Oliveira {et~al.}(2015)Oliveira, Velten, Fabris, \&
  Casarini}]{Oliveira:2015lka}
Oliveira, A.~M., Velten, H. E.~S., Fabris, J.~C., \& Casarini, L. 2015, Phys.
  Rev. D, 92, 044020, \dodoi{10.1103/PhysRevD.92.044020}

\bibitem[{Raaijmakers {et~al.}(2019)}]{Raaijmakers:2019qny}
Raaijmakers, G., {et~al.} 2019, Astrophys. J. Lett., 887,
  \dodoi{10.3847/2041-8213/ab451a}

\bibitem[{Raposo {et~al.}(2019)Raposo, Pani, Bezares, Palenzuela, \&
  Cardoso}]{Raposo:2018rjn}
Raposo, G., Pani, P., Bezares, M., Palenzuela, C., \& Cardoso, V. 2019, Phys.
  Rev. D, 99, 104072, \dodoi{10.1103/PhysRevD.99.104072}

\bibitem[{Rastall(1972)}]{Rastall:1972swe}
Rastall, P. 1972, Phys. Rev. D, 6, 3357, \dodoi{10.1103/PhysRevD.6.3357}

\bibitem[{Rastall(1976)}]{Rastall:1976uh}
---. 1976, Can. J. Phys., 54, 66, \dodoi{10.1139/p76-008}

\bibitem[{Reardon {et~al.}(2016)}]{Reardon:2015kba}
Reardon, D.~J., {et~al.} 2016, Mon. Not. Roy. Astron. Soc., 455, 1751,
  \dodoi{10.1093/mnras/stv2395}

\bibitem[{Reed \& Horowitz(2020)}]{Reed:2019ezm}
Reed, B., \& Horowitz, C.~J. 2020, Phys. Rev. C, 101, 045803,
  \dodoi{10.1103/PhysRevC.101.045803}

\bibitem[{Riley {et~al.}(2021)}]{Riley:2021pdl}
Riley, T.~E., {et~al.} 2021, Astrophys. J. Lett., 918,
  \dodoi{10.3847/2041-8213/ac0a81}

\bibitem[{Romani {et~al.}(2022)Romani, Kandel, Filippenko, Brink, \&
  Zheng}]{Romani:2022jhd}
Romani, R.~W., Kandel, D., Filippenko, A.~V., Brink, T.~G., \& Zheng, W. 2022,
  Astrophys. J. Lett., 934, \dodoi{10.3847/2041-8213/ac8007}

\bibitem[{Roupas \& Nashed(2020)}]{Roupas:2020mvs}
Roupas, Z., \& Nashed, G. G.~L. 2020, Eur. Phys. J. C, 80, 905,
  \dodoi{10.1140/epjc/s10052-020-08462-1}

\bibitem[{Tews {et~al.}(2018)Tews, Carlson, Gandolfi, \& Reddy}]{Tews:2018kmu}
Tews, I., Carlson, J., Gandolfi, S., \& Reddy, S. 2018, Astrophys. J., 860,
  149, \dodoi{10.3847/1538-4357/aac267}

\bibitem[{Traversi {et~al.}(2022)Traversi, Char, Pagliara, \&
  Drago}]{Traversi:2021fad}
Traversi, S., Char, P., Pagliara, G., \& Drago, A. 2022, Astron. Astrophys.,
  660, \dodoi{10.1051/0004-6361/202141544}

\bibitem[{Tsaloukidis {et~al.}(2022)Tsaloukidis, Koliogiannis, Kanakis-Pegios,
  \& Moustakidis}]{Tsaloukidis:2022rus}
Tsaloukidis, L., Koliogiannis, P.~S., Kanakis-Pegios, A., \& Moustakidis, C.~C.
  2022.
\newblock \doarXiv{2210.15644}

\bibitem[{Urbano \& Veerm\"ae(2019)}]{Urbano:2018nrs}
Urbano, A., \& Veerm\"ae, H. 2019, JCAP, 04, 011,
  \dodoi{10.1088/1475-7516/2019/04/011}

\bibitem[{Visser(2018)}]{Visser:2017gpz}
Visser, M. 2018, Phys. Lett. B, 782, 83, \dodoi{10.1016/j.physletb.2018.05.028}

\bibitem[{Witten(1984)}]{Witten:1984rs}
Witten, E. 1984, Phys. Rev. D, 30, 272, \dodoi{10.1103/PhysRevD.30.272}

\bibitem[{Yang {et~al.}(2020)Yang, Gayathri, Bartos, Haiman, Safarzadeh, \&
  Tagawa}]{Yang:2020xyi}
Yang, Y., Gayathri, V., Bartos, I., {et~al.} 2020, Astrophys. J. Lett., 901,
  \dodoi{10.3847/2041-8213/abb940}

\bibitem[{Zdunik {et~al.}(2001)Zdunik, Haensel, \& Gourgoulhon}]{Zdunik:2001yz}
Zdunik, J.~L., Haensel, P., \& Gourgoulhon, E. 2001, Astron. Astrophys., 372,
  535, \dodoi{10.1051/0004-6361:20010510}

\bibitem[{{Ziaie} {et~al.}(2020){Ziaie}, {Moradpour}, \&
  {Shabani}}]{2020EPJP..135..916Z}
{Ziaie}, A.~H., {Moradpour}, H., \& {Shabani}, H. 2020, European Physical
  Journal Plus, 135, 916, \dodoi{10.1140/epjp/s13360-020-00927-2}

\end{thebibliography}



\end{document}